\documentclass[a4paper,12pt]{article}
\usepackage{amsfonts}
\usepackage{graphicx}
\usepackage{amsmath}
\usepackage{color}
\usepackage{cite}

\newcommand{\be}{\begin{equation}}
\newcommand{\ee}{\end{equation}}
\begin{document}


\begin{center}
\noindent{{\LARGE{Comments on single trace $T\bar{T}$ and other current-current deformations}}}
\smallskip
\smallskip

\smallskip
\smallskip

\smallskip
\smallskip

\smallskip
\smallskip

\noindent{{Gaston Giribet$^{a}$, Julio Oliva$^{b}$, Ricardo Stuardo$^{b}$}}

\smallskip
\smallskip

\smallskip
\smallskip

\smallskip
\smallskip

\smallskip
\smallskip

\smallskip
\smallskip

$^a${Departamento de F\'{\i}sica, Universidad de Buenos Aires \& IFIBA - CONICET}\\ 
{\it Ciudad Universitaria, pabell\'on 1 (1428) Buenos Aires, Argentina.}

\smallskip

$^b${Departamento de F\'{i}sica, Universidad de Concepci\'on}\\
{\it Casilla 160-C, Concepci\'on, Chile.}


\end{center}

\smallskip
\smallskip

\smallskip
\smallskip

String theory on AdS$_3$ with NS-NS fluxes admits a solvable irrelevant deformation which is close to the $T\bar{T}$ deformation of the dual CFT$_2$. This consists of deforming the worldsheet action, namely the action of the $SL(2,\mathbb{R})$ WZW model, by adding to it the operator $J^-\bar{J}^-$, constructed with two Kac-Moody currents. The geometrical interpretation of the resulting theory is that of strings on a conformally flat background that interpolates between AdS$_3$ in the IR and a flat linear dilaton spacetime with Hagedorn spectrum in the UV, having passed through a transition region of positive curvature. Here, we study the properties of this string background both from the point of view of the low-energy effective theory and of the worldsheet CFT. We first study the geometrical properties of the semiclassical geometry, then we revise the computation of correlation functions and of the spectrum of the $J^-\bar{J}^-$-deformed worldsheet theory, and finally we discuss how to extend this type of current-current deformation to other conformal models.

\smallskip
\smallskip

\smallskip



\newpage

\section{Introduction}

In the context of AdS$_3$/CFT$_2$ correspondence, it was shown in \cite{Giveon:2017nie} that certain type of $T\bar{T}$-deformation of the boundary CFT$_2$, which can be regarded as a single trace version of the one originally introduced in \cite{Smirnov:2016lqw, Cavaglia:2016oda, McGough:2016lol}, gives rise in the bulk to a string theory background that interpolates between AdS$_3$ in the infrared limit and a flat linear dilaton background in the ultraviolet. This construction was argued in \cite{Giveon:2017nie} to provide a family of holographic pairs, including a large class of string theory vacua with asymptotically linear dilaton. The solvable irrelevant deformation of AdS$_3$/CFT$_2$ correspondence studied in \cite{Giveon:2017nie} was further studied in \cite{Giveon:2017myj}, where in particular its spectrum was studied. It was observed that this type of deformation leads in the ultraviolet to a theory with Hagedorn spectrum. This has been studied in \cite{Ben-Israel:2017zyi, Aharony:2018bad, Apolo:2019zai, Chakraborty:2020swe, Nuevo, Nuevoq} and references therein and thereof; see also \cite{ElUno, ElDos, ElTres}.

In \cite{Asrat:2017tzd, Giribet:2017imm}, the correlation functions in the deformed theory were studied, and it provided an alternative way of studying the spectrum: The insertion of the operator that realizes the deformation produces a logarithmic divergence in the correlation functions, leading to the renormalization of the primary operators. This yields an anomalous dimension that can be computed explicitly. From this, one may determine the spectrum of the theory from the worldsheet computation. The form of the correlation functions, on the other hand, permits to investigate the properties of the dual theory \cite{Asrat:2017tzd}.

The model studied in \cite{Giveon:2017nie} was later investigated in many different context. The entanglement entropy was first studied in \cite{Chakraborty:2018kpr}; in \cite{Babaro:2018cmq, Giribet:2020kde} the theory was studied in presence of boundaries; and the $J\bar{T}$ analog of it has also been studied \cite{Chakraborty:2019mdf, Apolo:2018qpq, Giveon:2019fgr, Apolo:2019yfj}. Here, we study the properties of this string background both from the point of view of the low-energy effective theory and of the worldsheet CFT. In section 2, we study the geometrical properties of the semiclassical geometry. We study the geometry as solution to the low-energy effective field theory, its T-dual background, the main properties of this specific deformation of AdS$_3$, and the field probes in such an spacetime. In section 3, we revise the computation of correlation functions of \cite{Asrat:2017tzd, Giribet:2017imm} and how it provides a direct way of studying the spectrum. Finally, in section 4 we discuss how to extend this type of deformation to other conformal models.

\section{Low energy theory}

\subsection{Interpolating background}

Let us start by considering the effective theory describing the low-energy limit of bosonic string theory. This is given by the field equations
\begin{align}
    &R_{\alpha \beta }+2 \nabla_{\alpha } \nabla_{\beta } \Phi-\frac{1}{4} H_{\alpha \mu \nu} H_{\beta }^{\mu \nu}=0, \label{Low1}\\ 
    &\nabla_{\alpha }\left(e^{-2 \Phi} H^{\alpha  \mu \nu}\right)=0,\label{Low2} \\ 
    &\nabla^{\alpha }\nabla_{\alpha } \Phi-2 \nabla_{\alpha } \Phi \nabla^{\alpha } \Phi+{2\alpha'}+\frac{1}{12} H_{\alpha \mu \nu} H^{\alpha \mu \nu}=0.\label{Low3}
\end{align}
where $H_{\mu \nu \rho }=\partial_{[\mu}B_{\nu \rho ]}$ is the field strengh associated to the Kalb-Ramond $B$-field, and $\Phi $ is the dilaton. These equations admit locally AdS$_3$ solutions \cite{HW},
    \begin{equation}
        ds^{2}=-\frac{r^{2}}{\ell^{2}} \, d t^{2}+\frac{\ell^{2}}{r^{2}}\, d r^{2}+r^{2} d \theta^{2},\label{La4}
    \end{equation}
provided the other backgrounds fields take the form
    \begin{equation}
        \Phi=\Phi_{0}, \ \ \  \ H_{\mu \nu \rho}=\frac{2r}{\ell }\, \epsilon_{\mu \nu \rho}.
    \end{equation}
The dilaton receives quantum (i.e. finite-$\alpha'$) corrections. Here, we will consider the convention $\alpha ' =1$, so that the semiclassical limit corresponds to large $k=\ell^{2}/\alpha '=\ell^2$.

As (\ref{La4}) describes the universal covering of AdS$_3$, we have $t\in \mathbb{R}$. The radial coordinate is $r\in \mathbb{R}_{\geq 0}$, with the boundary of the space being located at $r\rightarrow \infty $. If we take $\theta $ to be periodic with a period $2\pi$, the metric above corresponds to that of the massless BTZ geometry \cite{BTZ, BTZ2}. It will be convenient to consider coordinates $r=\ell e^{\phi}$ and $x=\ell \theta$. In these variables, the metric and the field strength take the form
    \begin{eqnarray}\label{DefMetric}
        d s^{2}=e^{2 \phi}\left(-dt^{2} + dx^{2}\right)+\ell^{2}d\phi^{2}, \ \ \  \ 
         H_{\mu\nu\rho } =\partial_{[\mu}B_{\nu \rho]}=2 e^{2 \phi} \epsilon_{\mu\nu\rho },
    \end{eqnarray}
where we now consider the covering $x\in \mathbb{R}$. That is, the non-vanishing component of the Kalb-Ramond field is $B_{x t}=e^{2 \phi}$ and grows when approaching the boundary at $\phi\rightarrow\infty$. 


Now, let us consider a deformation of (\ref{DefMetric}), given by   
    \begin{equation}\label{DefAdS11}
        ds^{2}=\frac{e^{2\phi}}{\lambda e^{2\phi}+1} (-dt^2+dx^2)  + \ell^{2} d\phi^{2},
    \end{equation}
with $\lambda$ being a real parameter. This metric solves the field equations (\ref{Low1}) for arbitrary $\lambda $ provided the Kalb-Ramond field and the dilaton are given by
    \begin{align}
        B_{x t}=\frac{2e^{2 \phi}}{\lambda e^{2 \phi}+1}\, , \ \ \ \
				\Phi = \Phi_0 -\phi  - \frac 12 \log (\lambda + e^{-2\phi}),
    \end{align}
respectively. Near the boundary, the dilaton becomes linear in $\phi $. We will be mostly interested in the case $\lambda \geq 0$, as for $\lambda <0 $ the geometry exhibits a singularity at $\phi=-\frac 12 \log | \lambda |$. In terms of the double null coordinates ${{u}}=(x+t)/\ell$ and $\bar{{u}}=(x-t)/\ell$, the fields take the following form
    \begin{equation}\label{DefAdS1}
    ds^{2}  = \ell^{2}d\phi^{2}+\frac{\ell^{2}d{{u}}\, d\bar{{{u}}}}{\lambda+e^{-2\phi}} \ , \ \     B=\frac{\ell^2 d{{u}} \wedge d\bar{{{u}}}}{\lambda+e^{-2\phi}}.
    \end{equation}

This solution has recently attracted much attention \cite{Giveon:2017nie, Giveon:2017myj, Apolo:2019zai, Chakraborty:2020swe, Nuevo, Nuevoq, Chakraborty:2018kpr, Asrat:2017tzd, Giribet:2017imm, Babaro:2018cmq, Giribet:2020kde, Chakraborty:2019mdf, Apolo:2018qpq, Giveon:2019fgr, Apolo:2019yfj} as it appears as an exact string background that corresponds to a marginal deformation of the worldsheet theory on AdS$_3\times \mathcal{N}$ which is closely related to the $T\bar{T}$-deformation of the dual CFT. 

Let us go back for a moment to the more familiar coordinates $r=\ell e^{\phi}$, namely
    \begin{equation}\label{DefAdS2}
        ds^{2} = -\frac{r^{2}}{r^{2}\lambda+\ell^2}dt^{2} + \frac{\ell^2 }{r^{2}} dr^{2} + \frac{r^{2}}{r^{2}\lambda+\ell^2}dx^{2},
    \end{equation}
in which it becomes evident that the geometry interpolates between AdS$_3$ and Minkowski space: While in the limit $r\ll \ell /\sqrt{\lambda}$ one recovers the metric (\ref{La4}), in the limit $r\gg \ell /\sqrt{\lambda}$ one gets $ds^2=-d\hat{t}^2+ d\hat{x}^2+d\hat{y}^2$ where $\hat{t}=t/\sqrt{\lambda}\ell$, $\hat{x}=x/\sqrt{\lambda}\ell$, $\hat{y}=\ell \phi $. The local isometry group of spacetime (\ref{DefAdS2}) for arbitrary value of $\lambda $ is $ISO(1,1)$ and is generated by the Killing vectors $\partial_{t}$, $\partial_{x}$ and $x\partial_{t} - t\partial_{x}$. It gets enhanced to the full $SL(2,\mathbb{R})\times SL(2,\mathbb{R})$ for $\lambda =0$ and to $ISO(2,1)$ in the limit $\lambda \to \infty $

The interpolating geometry (\ref{DefAdS1}) has very interesting properties. Apart from being fascinating in that it describes the transition between AdS$_3$ and the type of linear dilaton background  that appears in little string theory, geometry (\ref{DefAdS1}) exhibits peculiar features: {It admits a supersymmetric embedding in type IIB SUGRA, since it appears in the S-dual frame of the D1/D5 system \cite{Israel:2003ry}}. Besides, it is solvable in different limits: On the one hand, despite being a geometry of non-constant curvature, it turns out that the probe fields are integrable on it, and thus enables to gain intuition from the semiclassical analysis. On the other hand, the string worldsheet $\sigma $-model is an {\it exact} string solution; being a marginal deformation of a WZW model, can be solved explicitly in the sense that analytic expressions for the correlation functions can be obtained and the spectrum can be written down.

\subsection{T-duality}

Let us study some of the properties of (\ref{DefAdS1}), starting by noticing that it is dual to $pp$-waves on AdS$_3$. Spacetime \eqref{DefMetric} happens to be invariant under $t$- and $x$-translations, and so we can apply T-duality transformations along the direction generated by $\partial _t $ and $\partial _x$ in order to obtain new solutions $\tilde{g}_{\mu\nu},\tilde{B}_{\mu\nu},\tilde{\Phi}$ to the field equations \eqref{Low1}-\eqref{Low3}. At the level of the low-energy effective action, this amounts to applying the Buscher rules \cite{Buscher:1987qj, Buscher:1987sk}. For the $t$-direction, these transformation rules read
\begin{equation}
        \tilde{g}_{tt} = \frac{1}{g_{tt}},\ \ \ \  \tilde{g}_{ti} = \frac{B_{ti}}{g_{tt}},\ \ \ \
        \tilde{g}_{ij} = g_{ij} - \frac{g_{ti}g_{tj}}{g_{tt}}-\frac{B_{ti}B_{tj}}{g_{tt}},
\end{equation}
together with
\begin{equation}
        \tilde{B}_{ti} = \frac{g_{ti}}{g_{tt}} ,\ \ \ \
        \tilde{B}_{ij} = B_{ij} - 2\frac{g_{t[i}B_{j]t}}{g_{tt}}, \ \ \ \         \tilde{\Phi}   = \Phi - \frac{1}{2}\log\left(g_{tt}\right),
\end{equation}
where $i,j$ correspond to the coordinates other than $t$. After performing these transformations and renaming variables as $u=x$, $v=t$, and $y=\ell\phi $, one obtains
\begin{equation}\label{Tdual1}
        d\tilde{s}^{2} = -F(y)\, dv^{2} +2\, dudv+dy^{2}, \ \  \text{with}\ \ 
F(y)= \lambda + e^{-2{y}/{\ell}}
\end{equation}
together with $\tilde{B}_{\mu\nu}=0\, , \ \ \tilde{\Phi} = \tilde{\Phi}_{0} - \phi $. Analogously, applying similar transformations in the $x$-direction, one gets
\begin{equation}\label{Tdual2}
        d\tilde{s}^{2} = F(y)\, du^{2} +2\, dudv+dy^{2}\, .
\end{equation}
This geometry describes a 3-dimensional version of a $pp$-wave solution in Brinkmann type coordinates with a wave profile $F$. The non-vanishing component of the Riemann tensor for this geometry is 
\begin{equation}
R^{vy}_{\phantom{vy}uy} = -\frac{2}{l^2}e^{-2y/l}.
\end{equation}
This solution represents an {\it exact} string background.

\subsection{Geometric properties of the interpolating spacetime}

As said, spacetime \eqref{DefAdS1} interpolates between AdS$_3$ in the limit $\phi\rightarrow -\infty$ and a flat linear dilaton background in the opposite limit. The geometry thus has non-constant curvature. In fact, it can be shown to have an infinite region of positive curvature. To see this, we can compute the scalar curvature
                \begin{equation}
                    R = \frac{2(4\lambda r^{2}-3\ell^2)}{(\lambda r^{2}+\ell^2)^{2}},\label{LoR}
                \end{equation}
which, indeed, happens to be positive for $r> \ell \sqrt{3/(4\lambda )}$, all the way to infinity. $R$ has a global maximum at $r_{\text{max}} = \ell \sqrt{5/(2\lambda )}$ with a maximum value $R_{\text{max}} = 8/(7\ell^{2})$ that does not depend on the deformation parameter $\lambda $. Figure 1 depicts the function $R$ as a function of the radial coordinate $r$ for different values of $\lambda$.   
\begin{figure}[h]
    \begin{center}
        \includegraphics[width=8cm]{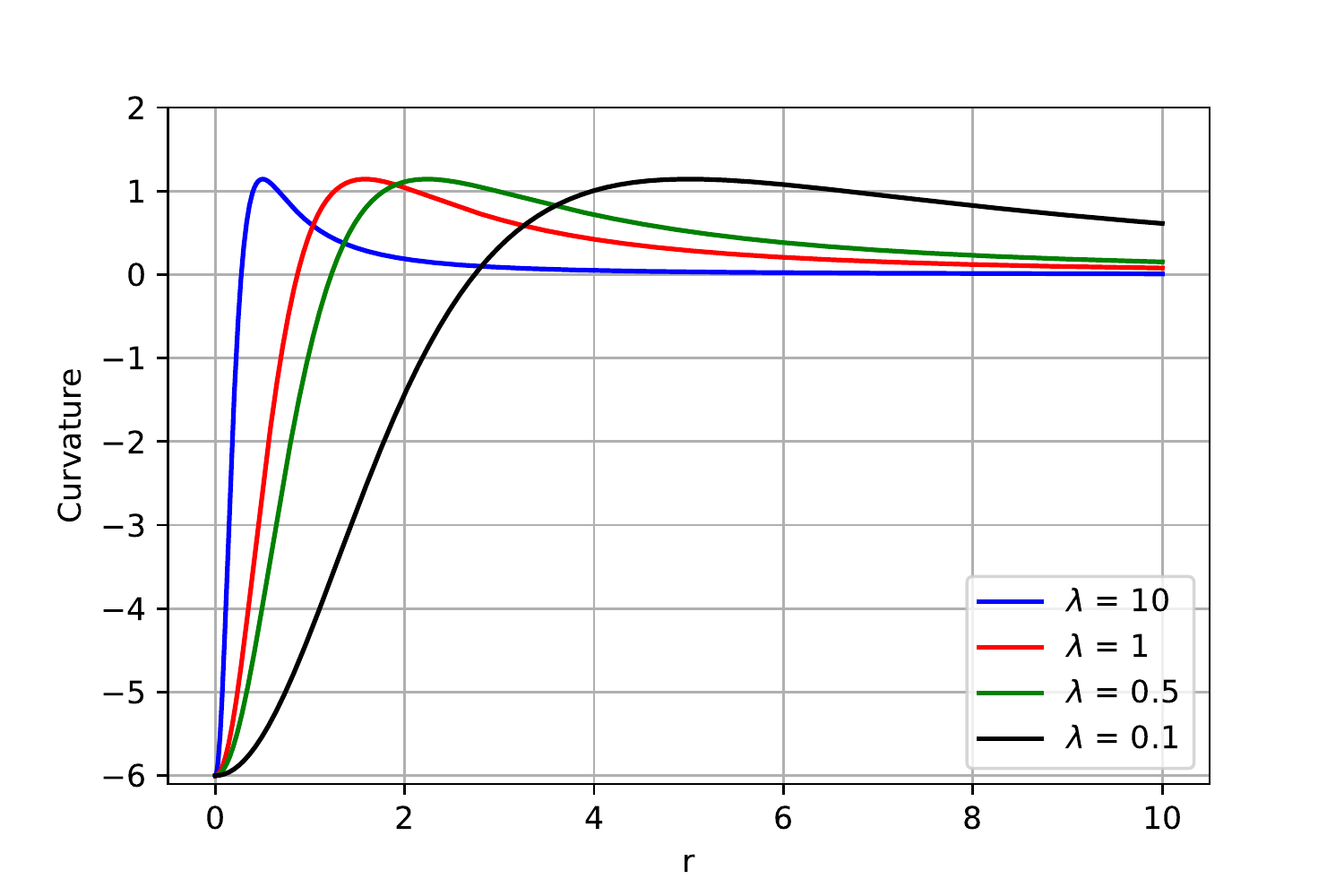}
        \caption{Ricci scalar for different values of $\lambda$ (here, $\ell = 1$).}
        \label{Ricci}
    \end{center}
\end{figure}

Other curvature invariants of (\ref{DefAdS1}) are
\begin{align}
&R^{\mu\nu}R_{\mu\nu} = \frac{ 4(6\lambda^{2}r^{4}-8l^{2}\lambda r^{2}+3l^{4} ) }{(\lambda r^{2}+l^{2})^{4}}, \\
R^{\mu}_{\phantom{\mu}\nu} &R^{\nu}_{\phantom{\nu}\rho}R^{\rho}_{\phantom{\rho}\mu} = \frac{-8(-10\lambda^{3}r^6+18l^{2}\lambda^{2}r^{4}-12l^{4}\lambda r^{2}+3l^{6})}{(\lambda r^{2}+l^{2})^{6}},
\end{align}
and we see from these, and from (\ref{LoR}), that the geometry is actually singular at $r= \ell /\sqrt{-\lambda}$. It is important noticing that it is sufficient to give the three curvature invariant $R$, $\text{Tr}(R_{\mu \nu}^2)$ and $\text{Tr}(R_{\mu \nu}^3)$ to characterize them all, since any higher curvature scalar can be obtained as a combination of powers of the latter three quantities by virtue of the three-dimensional identities
\begin{equation}
\delta^{\mu_1\cdots\mu_n}_{\nu_1\cdots\nu_n}\tilde R^{\nu_1}_{\ \mu_1}\cdots \tilde R^{\nu_n}_{\ \mu_n}\equiv 0 \ , n>3\ ,
\end{equation}
where $\tilde R^{\nu_1}_{\ \mu_1}$ is the traceless part of the Ricci tensor.
Despite being of non-constant curvature, geometry (\ref{DefAdS1}) yields vanishing Cotton tensor
\begin{equation}
C_{\mu \nu }= \epsilon_{\mu }^{\ \alpha \beta }\nabla_{\alpha } (R_{\nu \beta}-\frac 14 Rg_{\nu \beta })= 0
\end{equation}
which implies that it is locally conformally flat. This property makes Weyl invariant probes being integrable on this background, as we will show below. This permits to gain a semiclassical intuition. The conformal factor that allows to write the interpolating background \eqref{DefAdS2} in a manifestly conformally flat form, defines an improper Weyl transformations and, therefore, imposing boundary conditions and asymptotic behaviors on probe fields is non-trivially related to their flat counterpart.

\subsection{Probes on the deformed geometry}


Consider a conformally coupled scalar field on the geometry (\ref{DefAdS1}). The corresponding equation is
    \begin{equation}\label{CCSEQ}
         \partial_{\mu }(\sqrt{-g}g^{\mu\nu}\partial_{\nu })\Phi -  \frac{1}{8} {\sqrt{-g}} R\Phi=0.
    \end{equation}
We consider the separable ansatz
    \begin{equation}\label{La22}
        \Phi(t,r,x) = e^{-i\omega t}e^{i\kappa x}\varphi(r).
    \end{equation}
This problem is exactly solvable. However, we can first gain intuition from the well-known small and large $r$ regimes, where it reduces to the AdS$_3$ and to the flat space computation, respectively. This does not mean that the solution to the complete problem will be a simple junction of the two constant curvature problems. The transmission coefficients may actually change due to the different boundary conditions that have to be satisfied in the $\lambda $-deformed background.

Let us consider first the case $\kappa ^{2}-\omega^{2}<0$. In this case, the solution for $\varphi (r)$ takes the form
    \begin{equation}\label{sol}
        \begin{split}
       \varphi (r) &= \left(\frac{r^{2}\lambda+1}{r^{2}}\right)^{1/4}\left(
A_{1}\, e^{i\sqrt{\omega^{2}-\kappa^{2}}\, \chi (r)} + B_{1}\, e^{-i\sqrt{\omega^{2}-\kappa^{2}}\, \chi (r)} \right),
        \end{split}
    \end{equation}
where
\begin{equation}
\chi (r) = \sqrt{\lambda}\log\left(  \lambda r+\sqrt{(r^{2}\lambda+1)\lambda}   \right)-\frac{\sqrt{r^{2}\lambda + 1}}{r}  
\end{equation}
and where $A_1$ and $B_1$ are two constant to be determined by requiring appropriate boundary conditions. In order to impose conditions at infinity, it is convenient to solve \eqref{CCSEQ} on the nearly flat metric
    \begin{equation}\label{MetricAtInf}
        ds^{2} \simeq -\frac{1}{\lambda}dt^{2} + \frac{dr^{2}}{r^{2}} + \frac{1}{\lambda}dx^{2}
    \end{equation}
which is the large $r$ limit of \eqref{DefAdS2} (here, we set $\ell = 1$ for short). The radial dependence of the conformal scalar on this metric reads $ \varphi (r) \propto A_{1}\, e^{i\sqrt{\lambda}\sqrt{\omega^{2}-\kappa^{2}}\log(r)} + B_{2}\, e^{-i\sqrt{\lambda}\sqrt{\omega^{2}-\kappa^{2}}\log(r)}$. We want to impose outgoing boundary conditions at infinity. Since for $\lambda \neq 0$ we have
\begin{align}
    \Phi (t,r,x) &\sim A_{1}\, e^{-i\omega( t-\sqrt{\lambda}\log(r))} + B_{1}\, e^{-i\omega(t+\sqrt{\lambda}\log(r))} ,
\end{align}
by expanding at infinity we find that imposing outgoing boundary conditions corresponds to $B_{1}=0$. This is confirmed by considering the flux of particles defined by the $U(1)$-current $j^{\mu}=-i\left(\varphi^{*}\partial^{\mu}\varphi-\varphi\partial^{\mu}\varphi^{*}\right)$. Thus,
    \begin{equation}\label{sol2}
       \varphi (r) = A_{1}\left(\frac{r^{2}\lambda+1}{r^{2}}\right)^{1/4}
e^{i\sqrt{\omega^{2}-\kappa^{2}}\left(\sqrt{\lambda}\log\left(  \lambda r+\sqrt{(r^{2}\lambda+1)\lambda}   \right)-\frac{\sqrt{r^{2}\lambda + 1}}{r}  \right)}.
    \end{equation}    

Now, let us study the behavior near $r = 0$. To do so, we expand the expression around the origin, where we find that the dominant part goes like $\varphi (r) \sim {r^{-1/2}}{e^{-i{\sqrt{\omega^{2}-\kappa^{2}}}/{r}}}$. At this point, we are interested in making connection between this result and the well-known result for AdS$_3$ (i.e. $\lambda=0$) when $\omega^{2}>\kappa^{2}$; namely
    \begin{equation}\label{l0}
        \varphi _{\lambda=0}(z) = A_{2}\, z\, J_{\frac{1}{2}}\left(\sqrt{\omega^{2}-\kappa^{2}}z\right) + B_{2}\, z\, J_{-\frac{1}{2}}\left(\sqrt{\omega^{2}-\kappa^{2}}z\right)
    \end{equation}
where $z=1/r$ and where $J_{\pm 1/2}$ are Bessel functions\footnote{These Bessel functions actually reduce to elementary functions, which is due to the fact that on AdS our problem reduces to that of a scalar with the conformal mass $m^2\ell^2=m_{\text{conf}}^2\ell^2=-3/4$. For arbitrary values of the mass the Bessel functions are replaced by $J_{\pm\nu}$ with $\nu=\sqrt{1+m^2}$.}. In order to relate \eqref{l0} with the complex exponentials in (\ref{sol}) one can use
    \begin{equation}
       J_{\frac{1}{2}}(x) = \sqrt{\frac{2}{\pi}}\frac{\sin(x)}{\sqrt{x}} \ , \ \ \  
				J_{-\frac{1}{2}}(x) = \sqrt{\frac{2}{\pi}}\frac{\cos(x)}{\sqrt{x}}
    \end{equation}
and, therefore, in terms of $r$ this reads
    \begin{equation}\label{limR0}
        \varphi _{\lambda=0}(r) = (A_{2}+B_{2})\, \frac{e^{i\frac{\sqrt{\omega^{2}-\kappa^{2}}}{r}}}{r^{1/2}}-(A_{2}-B_{2})\, \frac{e^{-i\frac{\sqrt{\omega^{2}-\kappa^{2}}}{r}}}{r^{1/2}}\, .
    \end{equation}
We see that asymptotic behavior of \eqref{sol2} is a particular linear combination of the solutions in \eqref{limR0}, namely with $B_{2}=-A_{2}$. This is equivalent to setting very special mixing boundary conditions in the AdS$_3$ region $r\ll 1/\sqrt{\lambda }$. This makes it a difference with respect to the AdS$_3$ case $\lambda =0$.

Now, let us see what happens in the case $\kappa^{2}-\omega^{2}>0$, where the solution reads
    \begin{equation}\label{solkp}
       \varphi (r) = \left(\frac{r^{2}\lambda+1}{r^{2}}\right)^{1/4}\left(
C_{1}e^{-\sqrt{\kappa^{2}-\omega^{2}} \chi (r)}+  D_{1} e^{\sqrt{\kappa^{2}-\omega^{2}} \chi (r)} \right)
    \end{equation}
As the wave function is now confined, we want the solution to vanish at infinity; therefore, we set $D_{1}=0$. As before, one can first take a look at the solution in the case $\lambda=0$; namely
    \begin{equation}
        \varphi_{\lambda=0}(r) = \frac{C_{2}}{r}K_{\frac 12 }\left(\frac{\sqrt{\kappa^{2}-\omega^{2}}}{r}\right)+ 
        \frac{D_{2}}{r}I_{\frac 12 }\left(\frac{\sqrt{\kappa^{2}-\omega^{2}}}{r}\right);
    \end{equation}
and then rewrite the modified Bessel functions $K_{1/2}$, $I_{1/2}$ as 
    \begin{equation}
        K_{\frac 12 }(x)=\sqrt{\frac{\pi}{2}}\frac{e^{-x}}{\sqrt{x}}\  \ \ \ \
        I_{\frac 12 }(x)=\sqrt{\frac{2}{\pi}}\frac{\sinh(x)}{\sqrt{x}}
    \end{equation}
to make contact with the solution for $\lambda=0$,
    \begin{align}
        \varphi _{\lambda=0}(r) &= \frac{C_{2}}{r^{1/2}}e^{-\frac{\sqrt{\kappa ^{2}-\omega^{2}}}{r}}+ 
        \frac{D_{2}}{r^{1/2}}\sinh\left(\frac{\sqrt{\kappa ^{2}-\omega^{2}}}{r}\right). \label{solkpl0}
    \end{align}
Expanding the $\lambda \neq 0 $ solution near $r=0$, one obtains
    \begin{equation}\label{expsol2}
\varphi (r) = A_{1}e^{\frac{\sqrt{\kappa ^{2}-\omega^{2}}}{r}}(  r^{-1/2}\, {e^{-\frac{1}{2}\sqrt{k^{2}-\omega^{2}}\sqrt{\lambda}\log(\lambda)}}+\mathcal{O}(r^{1/2})   \, )\, ;
    \end{equation}
so, we see that \eqref{expsol2} corresponds to the linear combination $C_2=D_2$ above. Notice that, while \eqref{sol2} diverges as $\sim r^{-1/2}$ when $r$ tends to zero, \eqref{expsol2} is exponentially divergent in that limit. Condition $C_{2}=D_{2}$ also looks from the AdS$_3$ viewpoint ($r\ll 1/\sqrt{\lambda }$) as mixed boundary conditions which are induced by the presence of the unusual asymptotic of the deformed theory.

As the conformally coupled scalar, the Dirac action is also Weyl invariant;
therefore, it is natural to study a spin-$1/2$ probe on the deformed,
conformally flat background \eqref{DefAdS2}. Explicitly, the Dirac equation reads
    \begin{equation}
        \left( \gamma^{a}e_{a}^{\mu}\partial_{\mu}
        +\frac{1}{2}\omega_{\mu}^{ab}\gamma^{c}e_{c}^{\mu}J_{ab} \right) \Psi=0\, .
    \end{equation}
The spinor $\Psi$ will split in two components $\Psi^{T}=\left( \Psi_{1}
,\Psi_{2} \right)$. The defomed metric \eqref{DefAdS2} is of the form
    \begin{equation}
        ds^{2}= -f^{2}(r)dt^{2} + g^2(r)dr^{2} + f^2(r)  dx^{2}\, ,
    \end{equation}
for which we can choose dreibein and compute the spin connections, leading
respectively to
    \begin{equation}
        e^{0} = f\, dt \text{, } e^{1} = g\, dr \text{ and } e^{2} = f\, dx\, ,
    \end{equation}
    \begin{equation}
        \omega^{01} = \frac{f'}{g} dt
        \text{ and } \omega^{12} = -\frac{f^{\prime}}{g}dx\ .
    \end{equation}
It is useful to use the explicit, real representation of the Dirac matrices $      \gamma^{0}=i\sigma^{2},\gamma^{1}=\sigma^{1},\gamma^{2}=\sigma^{3}$. With these expressions at hand, Dirac equation leads to the coupled system
    \begin{align}
\frac{1}{f}\partial_{t}\Psi_{2}+\frac{1}{g}\partial_{r}\Psi_{2}+\frac{1}
{f}\partial_{x}\Psi_{1}+\frac{f^{\prime}}{fg} \Psi_{2} &  =0,\\
\frac{1}{f}\partial_{t}\Psi_{1}-\frac{1}{g}\partial_{r}\Psi_{1}+\frac{1}
{f}\partial_{x}\Psi_{2}-\frac{f^{\prime}}{fg} \Psi_{1} &  =0.
    \end{align}
Defining $\Psi_{i}\left(  t,r,x\right)  =e^{-i\omega t+i\kappa x}\psi_{i}\left(
r\right)  $ with $i=1,2$, we can integrate the radial profiles for the spinor
as
    \begin{align}
\psi_{1}\left(  r\right)=\left(  \frac{r^{2}\lambda+l^{2}}{r^{2}%
}\right)  ^{1/4}\varphi(r) \text{ and } \psi_{2}(r) = \frac{\omega}{\kappa}\psi_{1}(r) - \frac{i}{\kappa} \frac{1}{g(r)}\frac{d}{dr}\left( f(r)\psi_{1}(r) \right)\ ,
\end{align}
where $\varphi(r)$ is given by \eqref{sol} or \eqref{solkp} depending on the sign of $\kappa^2-\omega^2$, as before. Consequently, the asymptotic behavior for the fermion is inherited by that of the scalar. When the momentum along the direction $x$ vanishes, namely when $\kappa=0$ the integration for the Dirac field is simpler and leads to
\begin{align}
\psi_{1}(r)  & =C_{1}\left(  \frac{1+r^{2}\lambda}{r^{2}}\right)
^{1/2}\left(  \sqrt{\lambda}r+\sqrt{1+r^{2}\lambda}\right)  ^{-i\omega
\sqrt{\lambda}}e^{\frac{i\omega\sqrt{1+r^{2}\lambda}}{r}}\\
\psi_{2}(r)  & =C_{2}\left(  \frac{1+r^{2}\lambda}{r^{2}}\right)
^{1/2}\left(  \sqrt{\lambda}r+\sqrt{1+r^{2}\lambda}\right)  ^{i\omega
\sqrt{\lambda}}e^{-\frac{i\omega\sqrt{1+r^{2}\lambda}}{r}}%
\end{align}
The two independent solution $C_1=0$ or $C_2=0$, respectively describe an ingoing or outgoing flux of particles. Again, from the perspective of AdS$_3$, these would correspond to mixed boundary conditions.

\section{String theory} 

Now, let us study the string worldsheet theory. The worldsheet action on the background \eqref{DefAdS1} takes the form
    \begin{equation}\label{Polyakov}
            S = \frac{1}{2\pi} \int d^{2}z \left(\frac{1}{2}\partial\phi\bar{\partial}\phi 
						+\frac 12 \partial\bar{u}\bar{\partial}u (\lambda + e^{-\sqrt{2/(k-2)}\phi })^{-1} \right),
    \end{equation}
together with an extra linear dilaton term $-({1}/{2\pi}) \int d^2z\sqrt{{2}/({k-2})}{R}\phi $ with ${R}$ being here the worldsheet curvature. We see here that, provided $\lambda  \neq 0$, in the limit $\phi \to \infty $ one recovers the free theory with a background charge term. In the case $\lambda = 0$, in contrast, the theory at $\phi \to \infty $ exhibits a non-trivial coupling between $\phi$ and the $u, \bar{u}$ dependence. This can be regarded as an effective potential in the $\phi $-direction. This potential vanishes for holomorphic configurations such that $\partial\bar{u}\bar{\partial}u=0$. This configurations are closely related to the so-called long strings, which form a continuum in the spectrum in AdS$_3$.

\subsection{Strings on AdS$_3 \times \mathcal{N}$}

Let us begin by reviewing the undeformed theory ($\lambda  =0$), namely bosonic string theory on AdS$_3 \times \mathcal{N}$. This theory corresponds to the level-$k$ WZW model on $SL(2,\mathbb{R})$, and so it has $\hat{sl}(2)_k$ affine Kac-Moody symmetry, which is generated by local currents whose modes are usually denoted $J_n^{\pm }$, $J_n^{3}$, along with their anti-holomorphic counterparts. Virasoro symmetry follows from the Sugawara construction. We consider primary operators of the form 
    \begin{equation}
        V_{h}(p|z) = Z_0|p|^{2-2h}\, e^{ip{{u}}(z)+i\bar{p}\bar{{{u}}}(\bar{z})} e^{\sqrt{2/(k-2)}(h-1)\phi(z,\bar z )}\, \times \, ...
    \end{equation}
where the ellipsis stand for the contributions of internal part $\mathcal{N}$. These are the vertex operators of the theory. $p$ and $\bar{p}$ are the momenta conjugate to directions $u$ and $\bar{u}$, while $h$ is related to the radial momentum. The factor $Z_0|p|^{2-2h}$ stands for a normalization. The worldsheet conformal dimension of these operators are
\begin{eqnarray}
\Delta_{\lambda =0} = \frac{h(1-h)}{k-2} +\Delta_{\mathcal{N}}+ N \, \label{confo}
\end{eqnarray}
An analogous expression holds for $\bar \Delta _{\lambda =0}$ with $\bar \Delta_{\mathcal{N}}$ and $\bar N $. $\Delta_{\mathcal{N}}$ stand for the conformal dimension of the operators of the CFT on the internal space $\mathcal{N}$, and $N$ is the string excitation number. As just said, $p$ and $\bar p$ represent the momentum in the boundary, and they relate to the momentum in (\ref{La22}) as follows
\begin{equation}
\kappa = \frac{p + \bar{p}}{\ell } \ , \ \ \ \omega = \frac{ \bar{p} - p}{\ell }\, .
\end{equation}
In the Euclidean theory, $t\to i t$ and $\bar{p}$ is the complex conjugate of $p$. The index $h$ labels the representations of $SL(2,\mathbb{R})$. We focus on the long string states, which belong to the continuous series representations, having
\begin{equation}
h=\frac 12 + i s \, , \ \ \text{with} \ \ \ \  s\in \mathbb{R}.\label{arribeno}
\end{equation}
These long strings can reach the boundary due to the coupling to the $B$-field. They have a continuous energy spectrum, which depends on the spectral flow variable $w \in \mathbb{Z}_{\geq 0}$ that accounts for the winding number of the string around the boundary. To analyze the spectrum of the theory on AdS$_3\times \mathcal{N}$ in the momentum space, it is convenient to consider the operator basis
\begin{equation}
V_{h,m,\bar m}(z) = \frac{\Gamma(h+m)}{\Gamma(1-h-\bar m )} \int \frac{d^2p}{|p|^2}\, p^{-m}\bar{p}^{-\bar m }\, V_{h}(p|z)\, .
\end{equation}
Performing spectral flow transformation on the states created by these operators, one obtains the states of the sector $w$, whose conformal dimensions are
\begin{eqnarray}
\Delta_{\lambda =0} = \frac{h(1-h)}{k-2} -mw -\frac{k}{2}w^2+\Delta_{\mathcal{N}}+ N \, . 
\end{eqnarray}
where the energy is given by $m+\bar m + kw $ and the angular momentum by $m-\bar m $.

The 2-point function in the theory on AdS$_3 \times \mathcal{N}$ is well-known. For long strings in the basis $V_{h}(p|z)$, this takes the form 
    \begin{equation}\label{La43}
        \langle V_{\frac 12 +is_1}(p_1|z_{1}=0) V_{\frac 12 +is_2}(p_2|z_{2}=1)  \rangle_{\lambda=0}
        = \frac{2s_1}{\pi k} \, Z_0^2 \nu(k)^{2is_1}|p_1|^{4is_1}\, \delta^{(2)}(p_1+p_2)\, \delta(s_1-s_2)\, e^{2i\varphi }
				\end{equation}
where
\begin{equation}
   e^{2i\varphi } = {\Gamma}/{\Gamma^*} \, , \ \ \text{with} \ \ \ \ \Gamma = \Gamma(-2is) \Gamma({-2is}/({k-2})) 
\end{equation}
and
\begin{equation}
\nu(k)=\frac{\Gamma(\frac{1}{k-2})}{\Gamma(1- \frac{1}{k-2})}.
\end{equation}
The subscript ${\lambda =0}$ in (\ref{La43}) refers to the fact that the quantity corresponds to the undeformed AdS$_3\times \mathcal{N}$ background. In the deformed theory, the 2-point function has been computed in \cite{Asrat:2017tzd, Giribet:2017imm}, yielding
\begin{equation}
        \langle V_{h_1}(p_1|z_{1}=0) V_{h_2}(p_2|z_{2}=1)  \rangle_{\lambda }
        = \delta^{(2)}(p_1+p_2)\delta_{h_1-h_2}\, |p_1|^{4h_1-2} \, B(h_1)\label{kkkkk}
\end{equation}
with
\begin{equation}			\label{LLL}
				B(h_1)= \frac{\nu(k)^{2h_1-1}}{\pi} 
        \, 
            \frac{\Gamma(1-2h_1)\Gamma(1-\frac{2h_1-1}{k-2})}{\Gamma(2h_1-1)\Gamma(\frac{2h_1-1}{k-2})},
\end{equation}    
and where the spectrum of the theory is given by
\begin{equation}
h= \frac 12 \pm \frac 12 \sqrt{8(k-2)\lambda |p|^2-s^2}\, . \label{kkkk}
\end{equation}
This reduces to the 2-point function of the $SL(2,\mathbb{R})_k$ WZW model in the limit $\lambda=0$. Eq. (\ref{kkkk}) follows from imposing the Virasoro constraint $\Delta_{\lambda =0}=1$ on (\ref{confo}). 

\subsection{Turning on the deformation}

To see in detail how to obtain the correlator (\ref{kkkkk})-(\ref{LLL}) and (\ref{kkkk}), we may first rewrite action (\ref{Polyakov}) by adding auxiliary fields $v,\, \bar{v}$, yielding the equivalent action
    \begin{equation}\label{DefWZW}
            S_{\lambda } = \frac{1}{2\pi} \int d^{2}z \left(\frac{1}{2}\partial\phi\bar{\partial}\phi -{{v}}\bar{\partial}{{u}} - \bar{{{v}}}\partial\bar{{{u}}} - 2{{v}}\bar{{{v}}}e^{-\sqrt{2/(k-2)}\phi} 
                        -2\lambda {{v}}\bar{{{v}}}  \right),
    \end{equation}
where now the pair $(v,u)$ forms a commuting, dimension-(1,0) $(\beta, \gamma)$ ghost system. As we work in the conformal gauge, we are omitting here a background charge that represents the dilaton term. For $\lambda =0$, equation (\ref{DefWZW}) is, indeed, the WZW model written in Wakimoto variables \cite{Wakimoto:1986gf}. Nevertheless, we prefer to keep the notation $v,\, u$ to make contact with the spacetime interpretation (\ref{DefAdS1}). The action of the $\lambda$-deformed theory is thus given by $S_{SL(2,\mathbb{R}) \text{ WZW}}-2\lambda \int dz^2\, {{v}}\bar{{{v}}}$. This corresponds to a current-current deformation of the WZW model, with the deformation being realized by the operator $\lambda {{v}}\bar{{{v}}}$. This is consistent with the fact that the specific Kac-Moody current in these variables reads $J^-(z)= v(z) $. 

When trying to compute a correlation function such as $\langle V_{h_1}(p_1|z_{1}) V_{h_2}(p_2|z_{2})  \rangle_{\lambda}$ in the path integral approach, namely
\begin{equation}
\langle V_{h_1}(p_1|z_{1}) V_{h_2}(p_2|z_{2})  \rangle_{\lambda} = \int \mathcal{D}\phi\mathcal{D}^2u\mathcal{D}^2v \, e^{-S_{\lambda }}\, V_{h_1}(p_1|z_{1}) V_{h_2}(p_2|z_{2})\, ,
\end{equation}
the presence of the operator $\lambda \int d^2z\,v\bar{v}$ induces an ultraviolet divergent term in the effective action after integrating the fields $u,\bar{u}$ (see \cite{Giribet:2017imm} for more details). This makes the contribution of the correlators coming from the undeformed theory to factorize, and the deformation ends up contributing with an exponential that contains the conformal integral
    \begin{equation}
        I_{0} =\int d^{2}z \, |z-z_{1}|^{-2} |z-z_{2}|^{-2} .\label{La51}
    \end{equation}
This divergent integral appears frequently in quantum field theory calculations. For instance, it appears in the one-loop computation of the anomalous dimension of the composite operator $\bar{\psi}\psi$ in the Thirring model. The result of it is logarithmically divergent and it can be regularized using different methods. By introducing a regulator $\epsilon $, this can be resolved as
    \begin{equation}\label{La61}
        I_{\epsilon} = \left( 1+2\epsilon\log|z_{1}-z_{2}| +\mathcal{O}(\epsilon^{2}) \right)\left( \frac{2\pi}{\epsilon} + \mathcal{O}(\epsilon^{0}) \right),
    \end{equation}
and, after renormalizing the vertex operators by choosing $Z_{\epsilon} = e^{-2\lambda |p|^2/\epsilon }$, one obtains
\begin{equation}
       \langle V_{h_1}(p_1|z_{1}) V_{h_2}(p_2|z_{2})  \rangle_{\lambda } \sim \, |z_{1}-z_{2}|^{-4\Delta_{0} - 4 \lambda |p_1|^{2}}
\end{equation}
From this, it is possible to read the anomalous dimension induced by the deformation; namely
    \begin{equation}
        \Delta_{\lambda =0} \rightarrow \Delta _{{\lambda}}= \Delta_{\lambda = 0} + \lambda|p|^{2}.
    \end{equation}    
Finally, imposing the Virasoro constraint $\Delta _{{\lambda =1}}$ and writing it in terms of the quanties of the undeformed theory that satisfied $\Delta_{\lambda = 0}=1$, one gets (\ref{kkkk}), which reduces to (\ref{arribeno}) in the case $\lambda =0$. We observe that $h\in \mathbb{R}$ provided $-4|p|\sqrt{k\lambda }\geq s \geq +4|p|\sqrt{k\lambda }$; and $h\in \frac 12 + i\mathbb{R}$ provided $|s|>4|p| \sqrt{k\lambda_0}$. The overall factor $|p_1|^{4h_1-2}$ in the 2-point function and the dependence of $h_1$ on $\lambda $ has been studied in detail in \cite{Asrat:2017tzd} to investigate the properties of the dual theory, especially its non-locality encoded in a branch cut that the 2-point function of the $\lambda$-deformed theory exhibits.

\section{Generalizations}

The advantage of the computation of the anomalous dimension described above is that it admits a straightforward generalization to other models, such as the $SL(N,\mathbb{R})$ WZW models or their supersymmetric extensions. Despite not in all such cases one has a string $\sigma $-model interpretation of the CFT, this is still interesting from the CFT point of view as it provides a set of solvable non-rational models. The simplest extension of this sort is the $SL(N,\mathbb{R})$ WZW model. In that case, the action can in principle be written as a sum of a Gaussian piece and an interaction piece $S_I$; namely
\begin{equation}\label{torbellino}
    S_{SL(N,\mathbb{R})\text{ WZW}} = \frac{1}{2\pi}\int d^{2}z \left( \left( \partial \phi,\bar{\partial}\phi \right) 
                - \sum_{a=1}^{N(N-1)/2}({{{v}}}_a\bar{\partial}{{{u}}}_a +{\bar{{{v}}}}_a\partial {\bar{{{u}}}}_a) \right) + S_{I}.
\end{equation}
This involves a set of $N-1$ scalars and $N(N-1)$ copies of $\beta , \gamma $ systems, which here we keep denoting by $v_a , u_a$ with $a=1, 2, ... N(N-1)/2$. The scalars, $\phi _i$, with $i=1, 2, ... N-1$ form a vector in the space of roots of $sl(N)$. We denote $(\, .\, ,\, .\, )$ the product in this space of roots, which is defined in terms of the Cartan matrix $K_{ij}=(e_i,e_j)$ with $e_1,\, e_2, \, ..., \,e_{N-1}$ being the simple roots, with the $N-1$ fundamental weights $w_i$ satisfying $(w_i,e_j)=\delta_{ij}$. $\rho $ is the Weyl vector, i.e. the half-sum of all positive roots. The Lagrangian also includes a background charge term $\int d^2z\, (\rho,\phi){R}/\sqrt{k-N}$. 

As before, in the appropriate basis a solvable family of current-current deformation of the theory is given by the addition of the marginal operator
    \begin{equation}\label{La55}
     \sum^{N-1}_{i=1}  \frac{\lambda_{i}}{\pi} \int d^{2}z\, {{v}}_{i}\bar{{{v}}}_{i},
    \end{equation}
where the field $J^-_i(z)=v_i(z)$ with $i=1,2,...N-1$ correspond to the Abelian subalgebra formed by $N-1$ lowering operators. If we denote $H=(J^3_1, J^3_2, ... J^3_{N-1})$ the generators of the Cartan subalgebra, and $E=(J^+_1,J^+_2, ... J^+_{N(N-1)/2})$ and $F=(J^-_1,J^-_2, ... J^-_{N(N-1)/2})$ the raising and lowering operators, respectively, then there exists an ordering such that the first $N-1$ elements $F\supset (J^-_1,J^-_2, ... J^-_{N-1})$ form an Abelian subalgebra. More importantly, there exists a free field representation such that these $N-1$ fields are given by $J^-_i=v_i$ with $i=1,2,...N-1$. For the case $A_{N-1}$ with $N=2,3,4,5$ these representations have been explicitly constructed in the literature \cite{Awata:1991az, Awata:1991cv, Gerasimov:1990fi}, and the generic case has been extensively discussed \cite{Bershadsky:1989mf, Bouwknegt:1990wa, Bouwknegt:1991gf}. Let us first show how the argument goes for generic $N$ and then consider an illustrating particular case. 

Consider the operators
\begin{equation}\label{DDD}
V_h(p,z) = Z_0\, e^{\sqrt{2/(k-N)}(h,\phi (z) )} e^{i\sum_{a=1}^{{N(N-1)}/{2}}(p^au_a(z)+\bar{p}^a\bar{u}_a(z))}
\end{equation}
where $h=(h_1,h_2,...h_{N-1})$ is the vector of the space of roots, and $p=(p^1,p^2, ...p^{N(N-1)/2})$ are the momentum associated to the directions $u_a$; and consider the correlation functions
\begin{equation}
\left\langle V_{h}(p_1|z_1) V_{h}(p_2|z_2) \right\rangle_{\lambda_1 , ...\lambda_{N-1}} = 
\int \prod_{i=1}^{N-1}\mathcal{D}\phi_i \prod_{a=1}^{N(N-1)/2}\mathcal{D}^2u_a \mathcal{D}^2v_a \, e^{-S_{\lambda_1, ...,\lambda_{N-1}}} \, V_{h}(p_1|z_1) V_{h}(p_2|z_2)\, .
\end{equation}
After integrating in $u_i$ for some $i$ (those that correspond to the fields $u_i$ that do not appear other than in the kinetic term) the action (\ref{torbellino})-(\ref{La55}) being linear in these fields, one obtains
\begin{equation}
        \bar{\partial}{{v}}_{i} = 2\pi i ( p^i_{1} \delta^{2}(z-z_{1}) + p^i_{2} \delta^{2}(z-z_{2})),
    \end{equation}
The solution is
    \begin{equation}\label{beta}
        {{v}}_{i}(z) =  \frac{ip^{i}_{1}}{z-z_{1}}- \frac{ip^{i}_{1}}{z-z_{2}}, 
    \end{equation}    
where we have used that, on the sphere, $p_{1}+p_{2}=0$ in virtue of the Riemann-Roch theorem. This can now be inserted back in (\ref{La55}). When doing so, one observes that a logarithmically divergent integral similar to (\ref{La51}) appears, yielding an anomalous correction to the conformal dimension of operators (\ref{DDD}). To see this in detail, let us consider the case $N=3$, in which the undeformed theory is given by the WZW model on $SL(3,\mathbb{R})$, whose action  reads
\begin{equation}
\begin{aligned}
    S_{SL(3,\mathbb{R})\text{ WZW}}=\frac{1}{2\pi}\int d^{2}z  &\left(\left( \partial \phi, \bar{\partial}\phi \right)         -\sum_{a=1}^{3} ({{v}}_{a}\bar{\partial}{{u}}_{a}
		+{\bar{v}}_{a}{\partial}{\bar{u}}_{a})+\right. \\
            & \left| {{v}}_{2} + {{v}}_{1}{{u}}_{3} \right|^{2}e^{\sqrt{2/(k-3)}\left(e_{2},\phi\right)} - {{v}}_{3}\bar{{{v}}}_{3}e^{\sqrt{2/(k-3)}\left(e_{3},\phi\right)} -\\
            &\left. {{v}}_{1}\bar{{{v}}}_{1}e^{\sqrt{2/(k-3)}\left(\rho,\phi\right)} 
             \right).
\end{aligned}
\end{equation}
together with a background charge term $\int d^2z\ ( \rho,\phi )R/\sqrt{k-3}$. As before, this action can be written in terms of the Wakimoto variables in such a way that two commuting currents take a simple form 
$J^{-}_{1}(z) = {{v}}_{1}(z)$ and $J^{-}_{2}(z) = {{v}}_{2}(z)$. Therefore, in the spirit of the deformation for $SL(2,\mathbb{R})$, we deform the
$SL(3,\mathbb{R})$ WZW model by adding to it two quadratic operators, for ${{v}}_{1}$ and ${{v}}_{2}$; namely
    \begin{equation}
S_{\lambda_1, \lambda_2} =       S_{SL(3,\mathbb{R})\text{ WZW}} - \frac{\lambda_{1}}{\pi}\int d^{2}z\, {{v}}_{1}\bar{{{v}}}_{1} 
                    - \frac{\lambda_{2}}{\pi}\int d^{2}z\, {{v}}_{2}\bar{{{v}}}_{2}.
    \end{equation}
We consider operators (\ref{DDD}) with $N=3$, $h=(h_1, h_2)$ and $p=(p_1,p_2, p_3)$, and the correlation function $\left\langle V_{h}( p,z_{1})V_{h}(-p,z_{2}) \right\rangle_{\lambda_1 , \lambda_2}$. After integrating on $u_i$, one finds the solutions for ${{v}}_{1}$ and ${{v}}_{2}$ to be
    \begin{equation}
        {{v}}_{1}(z) = \frac{ ip^{1}(z_{1}-z_{2}) }{(z-z_{1})(z-z_{2})}\ , \ \ \ 
        {{v}}_{2}(z) = \frac{ ip^{2}(z_{1}-z_{2}) }{(z-z_{1})(z-z_{2})}.
    \end{equation}
which, when replaced in the action, yields
\begin{equation}\label{La73}
\begin{aligned}
    S_{\lambda_1 , \lambda_ 2}=\frac{1}{2\pi}\int d^{2}z  &\left(\left( \partial \phi, \bar{\partial}\phi \right)  -{{v}}_{3}\bar{\partial}{{u}}_{3}- \bar{{{v}}_{3}}\partial\bar{{{u}}}_{3}
             -{{v}}_{3}\bar{{{v}}}_{3}e^{(\phi_{2} - \sqrt{3}\phi_{3})/\sqrt{k-3}} \right.   \\
            &\left. + \left|p^{2} + p^{1}{{u}}_{3}\right|^{2} e^{(\phi_{2} + \sqrt{3}\phi_{3})/\sqrt{k-3}} -|p^{1}|^{2}e^{2\phi_{2}/\sqrt{k-3}} \right) \\
            &+ \frac{1}{\pi} (\lambda_{1}|p^1|^2+\lambda_2 |p^2|^2)\, |z_1-z_2|^2\, I_0,
\end{aligned}
\end{equation}
with $I_{0}$ given by \eqref{La51}. Regularizing as in (\ref{La61}) and renormalizing the vertices accordingly, one obtains the corrected conformal dimension  
    \begin{equation}
         \Delta_{\lambda_1,\lambda_2} = \Delta_{\lambda_1=\lambda_2=0} + \lambda_{1}\left|p^{1}\right|^{2}+ \lambda_{2}\left|p^{2}\right|^{2}.
    \end{equation}
This follows from the last line in (\ref{La73}), which contains the logarithmic dependence in (\ref{La61}). This manifestly shows that the method of \cite{Giribet:2017imm} can be straightforwardly adapted to higher-rank.

\[
\]

\textbf{Acknowledgments}

G.G. thanks Edmundo Lavia and Mat\'{\i}as Leoni for discussions and collaborations. This work has been partially supported by CONICET through the grant PIP 1109-2017 and by Proyecto de Cooperación Internacional 2019/13231-7 FAPESP/ANID. The work of R.S. is funded by ANID Fellowships 22191591. The work of J.O. partially funded by FONDECYT grant 1181047.


\begin{thebibliography}{99}


\bibitem{Giveon:2017nie} 
  A.~Giveon, N.~Itzhaki and D.~Kutasov,
  ``$ \mathrm{T}\overline{\mathrm{T}} $ and LST,''
  JHEP {\bf 1707}, 122 (2017)
  [arXiv:1701.05576 [hep-th]].






\bibitem{Smirnov:2016lqw} 
  F.~A.~Smirnov and A.~B.~Zamolodchikov,
  ``On space of integrable quantum field theories,''
  Nucl.\ Phys.\ B {\bf 915}, 363 (2017)
  [arXiv:1608.05499 [hep-th]].

\bibitem{Cavaglia:2016oda} 
  A.~Cavagli\`a, S.~Negro, I.~M.~Sz\'ecs\'enyi and R.~Tateo,
  ``$T \bar{T}$-deformed 2D Quantum Field Theories,''
  JHEP {\bf 1610}, 112 (2016)
  [arXiv:1608.05534 [hep-th]].




\bibitem{McGough:2016lol}
L.~McGough, M.~Mezei and H.~Verlinde,
``Moving the CFT into the bulk with $ T\overline{T} $,''
JHEP \textbf{04}, 010 (2018)
[arXiv:1611.03470 [hep-th]].




\bibitem{Giveon:2017myj} 
  A.~Giveon, N.~Itzhaki and D.~Kutasov,
  ``A solvable irrelevant deformation of AdS$_{3}$/CFT$_{2}$,''
  JHEP {\bf 1712}, 155 (2017)
  [arXiv:1707.05800 [hep-th]].




\bibitem{Ben-Israel:2017zyi} 
  R.~Ben-Israel, A.~Giveon, N.~Itzhaki and L.~Liram,
  ``On the black hole interior in string theory,''
  JHEP {\bf 1705}, 094 (2017)
  [arXiv:1702.03583 [hep-th]].
  
\bibitem{Aharony:2018bad} 
  O.~Aharony, S.~Datta, A.~Giveon, Y.~Jiang and D.~Kutasov,
  ``Modular invariance and uniqueness of $T\bar{T}$ deformed CFT,''
  JHEP {\bf 1901}, 086 (2019)
  [arXiv:1808.02492 [hep-th]].
  
\bibitem{Apolo:2019zai}
L.~Apolo, S.~Detournay and W.~Song,
``TsT, $T\bar{T}$ and black strings,''
JHEP \textbf{06}, 109 (2020)
[arXiv:1911.12359 [hep-th]].



\bibitem{Chakraborty:2020swe}
S.~Chakraborty, A.~Giveon and D.~Kutasov,
``$T\bar T$, Black Holes and Negative Strings,''
[arXiv:2006.13249 [hep-th]].


\bibitem{Nuevo} 
S. Chakraborty, A. Giveon, D. Kutasov
``Comments on D3-Brane Holography,''
[arXiv:2006.14129 [hrp-th]].

\bibitem{Nuevoq}
S.~Chakraborty, A.~Giveon and D.~Kutasov,
``Strings in Irrelevant Deformations of $AdS_3/CFT_2$,''
[arXiv:2009.03929 [hep-th]].


\bibitem{ElUno}
S.~Chakraborty,
``$\frac{SL(2,\mathbb{R})\times U(1)}{U(1)}$ CFT, NS5$+$F1 system and single trace $T\bar{T}$,''
[arXiv:2012.03995 [hep-th]].

\bibitem{ElDos}
T.~Araujo, E.~\'O.~Colg\'ain, Y.~Sakatani, M.~M.~Sheikh-Jabbari and H.~Yavartanoo,
``Holographic integration of $T \bar{T}$ \textbackslash{}\& $J \bar{T}$ via $O(d,d)$,''
JHEP \textbf{03}, 168 (2019)
[arXiv:1811.03050 [hep-th]].

\bibitem{ElTres}
J.~L.~F.~Barbon and E.~Rabinovici,
``Remarks on the thermodynamic stability of $T \bar T$ deformations,''
J. Phys. A \textbf{53}, no.42, 424001 (2020)
[arXiv:2004.10138 [hep-th]].






\bibitem{Asrat:2017tzd} 
  M.~Asrat, A.~Giveon, N.~Itzhaki and D.~Kutasov,
  ``Holography Beyond AdS,''
  Nucl.\ Phys.\ B {\bf 932}, 241 (2018)
  [arXiv:1711.02690 [hep-th]].


\bibitem{Giribet:2017imm} 
  G.~Giribet,
  ``$T\bar{T}$-deformations, AdS/CFT and correlation functions,''
  JHEP {\bf 1802}, 114 (2018)
  [arXiv:1711.02716 [hep-th]].


\bibitem{Chakraborty:2018kpr} 
  S.~Chakraborty, A.~Giveon, N.~Itzhaki and D.~Kutasov,
  ``Entanglement beyond AdS,''
  Nucl.\ Phys.\ B {\bf 935}, 290 (2018)
  [arXiv:1805.06286 [hep-th]].




  
\bibitem{Babaro:2018cmq} 
  J.~P.~Babaro, V.~F.~Foit, G.~Giribet and M.~Leoni,
  ``$ T\overline{T} $ type deformation in the presence of a boundary,''
  JHEP {\bf 1808}, 096 (2018)
  [arXiv:1806.10713 [hep-th]].

\bibitem{Giribet:2020kde}
G.~Giribet and M.~Leoni,
``Current-current deformations, conformal integrals and correlation functions,''
JHEP \textbf{04}, 194 (2020)
[arXiv:2003.02864 [hep-th]].




\bibitem{Chakraborty:2019mdf} 
  S.~Chakraborty, A.~Giveon and D.~Kutasov,
  ``$T\bar{T}$, $J\bar{T}$, $T\bar{J}$ and String Theory,''
  J.\ Phys.\ A {\bf 52}, no. 38, 384003 (2019)
  [arXiv:1905.00051 [hep-th]].

\bibitem{Apolo:2018qpq}
L.~Apolo and W.~Song,
``Strings on warped AdS$_{3}$ via $ \mathrm{T}\bar{\mathrm{J}} $ deformations,''
JHEP \textbf{10}, 165 (2018)
[arXiv:1806.10127 [hep-th]].


\bibitem{Giveon:2019fgr}
A.~Giveon,
``Comments on $T\bar T$, $J\bar{T}$ and String Theory,''
[arXiv:1903.06883 [hep-th]].
  

\bibitem{Apolo:2019yfj}
L.~Apolo and W.~Song,
``Heating up holography for single-trace $J\bar{T}$ deformations,''
JHEP \textbf{01}, 141 (2020)
[arXiv:1907.03745 [hep-th]].



\bibitem{HW}
G.~T.~Horowitz and D.~L.~Welch,
``Exact three-dimensional black holes in string theory,''
Phys. Rev. Lett. \textbf{71}, 328-331 (1993)
[arXiv:hep-th/9302126 [hep-th]].

\bibitem{BTZ}
M.~Banados, C.~Teitelboim and J.~Zanelli,
``The Black hole in three-dimensional space-time,''
Phys. Rev. Lett. \textbf{69}, 1849-1851 (1992)
[arXiv:hep-th/9204099 [hep-th]].

\bibitem{BTZ2}
M.~Banados, M.~Henneaux, C.~Teitelboim and J.~Zanelli,
``Geometry of the (2+1) black hole,''
Phys. Rev. D \textbf{48}, 1506-1525 (1993)
[erratum: Phys. Rev. D \textbf{88}, 069902 (2013)]
[arXiv:gr-qc/9302012 [gr-qc]].

\bibitem{Israel:2003ry}
D.~Israel, C.~Kounnas and M.~P.~Petropoulos,
JHEP \textbf{10}, 028 (2003)
doi:10.1088/1126-6708/2003/10/028
[arXiv:hep-th/0306053 [hep-th]].

\bibitem{Buscher:1987qj}
T.~Buscher,
``Path Integral Derivation of Quantum Duality in Nonlinear Sigma Models,''
Phys. Lett. B \textbf{201}, 466-472 (1988)

\bibitem{Buscher:1987sk}
T.~Buscher,
``A Symmetry of the String Background Field Equations,''
Phys. Lett. B \textbf{194}, 59-62 (1987)




\bibitem{Wakimoto:1986gf} 
  M.~Wakimoto,
  ``Fock representations of the affine lie algebra A1(1),''
  Commun.\ Math.\ Phys.\  {\bf 104}, 605 (1986).
  
\bibitem{Awata:1991az} 
  H.~Awata, A.~Tsuchiya and Y.~Yamada,
  ``Integral formulas for the WZNW correlation functions,''
  Nucl.\ Phys.\ B {\bf 365}, 680 (1991).
  
\bibitem{Awata:1991cv} 
  H.~Awata,
  ``Screening currents ward identity and integral formulas for the WZNW correlation functions,''
  Prog.\ Theor.\ Phys.\ Suppl.\  {\bf 110}, 303 (1992)
  [hep-th/9202032].
  
\bibitem{Gerasimov:1990fi} 
  A.~Gerasimov, A.~Morozov, M.~Olshanetsky, A.~Marshakov and S.~L.~Shatashvili,
  ``Wess-Zumino-Witten model as a theory of free fields,''
  Int.\ J.\ Mod.\ Phys.\ A {\bf 5}, 2495 (1990).
  
\bibitem{Bershadsky:1989mf}
M.~Bershadsky and H.~Ooguri,
``Hidden SL(n) Symmetry in Conformal Field Theories,''
Commun. Math. Phys. \textbf{126}, 49 (1989)
  
\bibitem{Bouwknegt:1990wa}
P.~Bouwknegt, J.~G.~McCarthy and K.~Pilch,
``Free field approach to two-dimensional conformal field theories,''
Prog. Theor. Phys. Suppl. \textbf{102}, 67-135 (1990)

\bibitem{Bouwknegt:1991gf}
P.~Bouwknegt, J.~G.~McCarthy and K.~Pilch,
``Some aspects of free field resolutions in 2-D CFT with application to the quantum Drinfeld-Sokolov reduction,''
[arXiv:hep-th/9110007 [hep-th]].





  

\end{thebibliography}
\end{document}